\documentclass[
prb,twocolumn,
]{revtex4-2}

\usepackage{graphicx}
\usepackage{dcolumn}
\usepackage{bm}
\usepackage[svgnames]{xcolor}
\usepackage{mathtools}
\usepackage{dsfont}
\usepackage{url}
\usepackage{hyperref}
\usepackage[version=4]{mhchem}
\usepackage{braket}

\begin{document}

\preprint{APS/123-QED}

\title{Charged magnons on the surface of a topological Insulator}

\author{I. Martinez-Berumen}
\email{pablo.martinez2@mail.mcgill.ca}
\author{W. A. Coish}%
\email{coish@physics.mcgill.ca}
\author{T. Pereg-Barnea}
\email{tamipb@physics.mcgill.ca}

\affiliation{%
    Department of Physics, McGill University, Montr\'eal, Qu\'ebec, Canada H3A 2T8
}%

\date{\today}

\begin{abstract}
We study a system of two-dimensional Dirac electrons (as is realized on the surface of a 3D topological insulator) coupled to an array of localized spins. The spins are coupled ferromagnetically to each other, forming an ordered ground state with low-energy spin-wave excitations (magnons). The Dirac electrons couple to the spins through a spin-dependent effective Zeeman field. The out-of-plane effective Zeeman field therefore serves as a Dirac mass that gaps the electronic spectrum. Once a spin is flipped, it creates a surrounding domain in which the sign of the Dirac mass is opposite to that of the rest of the sample.  Therefore, an electronic bound state appears on the domain wall, as predicted by Jackiw and Rebbi. However, in a quantum magnet, a localized spin flip does not produce an eigenstate. Instead, the eigenstates correspond to delocalized spin waves (magnons).  As in the case of the single flipped spin, the delocalized magnon also binds an in-gap electronic state. We name this excitation a `Jackiw-Rebbi-Magnon' (JRM) and study its signature in the dynamic spin susceptibility. When the sample is tunnel-coupled to an electronic reservoir, a magnon produced in a system without any electrons hybridizes with a JRM (which binds a single electron), producing magnon-JRM polaritons. For such a system, we identify a quantum phase transition when the magnon-JRM polariton energy falls below that of the fully polarized ferromagnetic ground state. 

\end{abstract}

\maketitle


\section{Introduction}

The spectrum of a strong 3D topological insulator (3DTI) is gapped in the bulk and exhibits gapless surface states described by a Dirac cone, protected by time-reversal symmetry \cite{hasan2010colloquium,qi2011topological,fu2007topological}.
The addition of magnetic moments on the surface breaks time-reversal symmetry; these magnetic moments produce a local effective Zeeman field for the Dirac electrons rendering them massive \cite{liu2009magnetic,yu2010quantized,Tokura2019}. Such moments can be realized as magnetic impurities/dopants in topological insulators, producing magnetic topological insulators \cite{Lee2015,Chen2010,Xu2012,chang2013experimental}. Equivalently, these magnetic moments can be introduced as a separate system in proximity to the sample surface \cite{eremeev2013magnetic,vobornik2011magnetic,watanabe2019quantum}. The surface magnetic moments may interact with each other directly or through the  Ruderman–Kittel–Kasuya–Yosida (RKKY) interaction mediated by the Dirac electrons on the surface.  It has been shown that the associated RKKY interaction is likely to be ferromagnetic \cite{liu2009magnetic,abanin2011ordering}. When the moments are ordered ferromagnetically, the electrons acquire a finite Dirac mass everywhere on the surface \cite{liu2009magnetic,yu2010quantized}, resulting in a spectral gap. However, when the ferromagnet exhibits disorder (e.g., domains or skyrmions), the Dirac mass may change sign, resulting in the closure of the spectral gap \cite{hurst2015charged,Kunal2017magnetoconductance,liu2009magnetic}.

In the presence of magnetic domain walls where the Dirac mass changes sign, in-gap chiral bound states known as Jackiw-Rebbi modes appear \cite{jackiw1976solitons}. These states are exponentially localized to the domain wall as no extended states can exist at an energy within the surface gap \cite{hasan2010colloquium,liu2009magnetic}. While localized to domain walls, the Jackiw-Rebbi states may carry currents along the domain walls \cite{yasuda2017quantized}. Domain-wall chiral currents have been seen to contribute to experimental measurements such as the hysteretic magnetoconductance in magnetic topological insulator surfaces \cite{yasuda2017quantized,nakajima2016one,checkelsky2012dirac,rosen2017chiral,wang2016visualizing}. Theoretical models have also been proposed to study the formation and dynamics of domain walls in an applied magnetic field  \cite{Kunal2017magnetoconductance}.

\begin{figure}
    \centering
    \includegraphics{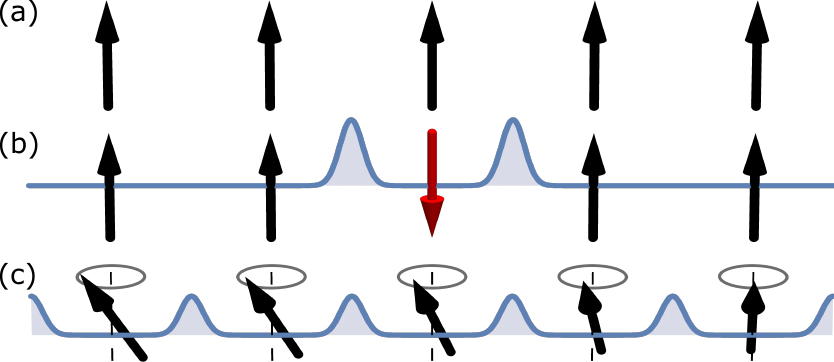}
    \caption{(a) Magnetic impurities in the ferromagnetic ground state in a magnetic topological insulator do not allow for electronic bound states. (b) A local mass change produced by an impurity spin flip allows for in-gap Jackiw-Rebbi modes. (c) A spin wave produces a delocalized domain wall with attached delocalized bound states.}
    \label{fig:fig1}
\end{figure}

In this paper, we consider a 3D topological insulator having surface electrons that are coupled to an array of localized spins. This coupling may arise, e.g., from exchange or from a local Zeeman interaction. Previous studies of 3D topological insulators coupled to insulating magnets have described the magnetic system classically \cite{Kunal2017magnetoconductance,efimkin2021topological,tserkovnyak2012thin}. In contrast, in this work we consider a quantum ferromagnet having low-energy spin-wave (magnon) excitations. These excitations can be considered as delocalized domain walls or as a superposition of domains in different configurations. The consequence, as we show, is the formation of delocalized chiral bound states that are attached to the magnon and we therefore name them `Jackiw-Rebbi Magnons' (see Fig.~\ref{fig:fig1}).

The rest of this paper is organized as follows: In Sec.~\ref{sec:model} we introduce the model for the surface states, the magnetic impurities, and the coupling between them. In Sec.~\ref{sec:jackiw_rebbi_magnons} we present the form of the elementary excitations, the Jackiw-Rebbi Magnons (JRMs) and, as an example, we numerically solve a lattice model having an appropriate long-wavelength limit. In Sec.~\ref{sec:spin_susceptibility} we propose and calculate an experimental observable, the dynamic transverse spin susceptibility, that can be used to probe JRM excitations. Finally, in Sec.~\ref{sec:Conclusions} we summarize our results and discuss the limitations and scope of the model.  

\section{Model}\label{sec:model}
We consider a model of ferromagnetically coupled magnetic impurities interacting with itinerant electrons on the surface of a 3D topological insulator. The system Hamiltonian is
\begin{equation}
    H_S=H_m+H_{em},
    \label{eq:H_complete}
\end{equation}
where $H_m$ is the Hamiltonian for the magnetic system and where $H_{em}$ describes the electrons in contact with the magnetic impurities.

The magnetic system is described by the (ferromagnetic) Heisenberg Hamiltonian (setting $\hbar=1$):
\begin{equation}
    H_m = 
    -J\sum_{\braket{ij}}
    \mathbf{S}_i \cdot \mathbf{S}_j
    -
    b \sum_{i} S^z_i,
\label{eq:HeisenbergHam}
\end{equation}
where $J>0$ is the exchange interaction between nearest neighbors, $\mathbf{S}_i$ is the spin operator for the impurity at site $i$, and $b$ is the effective Zeeman splitting. 
We assume that the magnetic impurities form a Bravais lattice, not necessarily commensurate with the underlying crystalline lattice. In this equation, and in the rest of this paper, we set $\hbar=1$, but $\hbar$ will be restored in several formulas below for clarity.  

In the absence of the electronic system, the Heisenberg Hamiltonian $H_m$ has a fully spin-polarized ferromagnetic ground state. If, without loss of generality, we assume $b>0$, this ground state is
\begin{equation}
    \ket{\mathrm{FM}}\equiv\ket{S,S,\hdots,S},
    \label{eq:FMstate}
\end{equation}
where $\ket{m_1,m_2,\hdots,m_{N}}$ is a simultaneous  eigenstate of all operators $\{S^z_i\}$ ($i=1,2,\ldots,N$) with eigenvalues $\{m_i\}$ ($S$ is the total impurity spin such that $|m_i|\le S$).
The energy of the ferromagnetic ground state is
\begin{equation}\label{eq:E_FM}
    E_{\mathrm{FM}}=-\frac{J}{2}NzS^2-bNS,
\end{equation}
where $z$ is the coordination number.
The low-energy excitations are the spin waves (magnons) $\ket{\mathbf{q}}$: $H_m\ket{\mathbf{q}}=E_\mathbf{q}\ket{\mathbf{q}}$, with
\begin{equation}
    \ket{\mathbf{q}}=\frac{1}{\sqrt{2S}}S_{\mathbf{q}}^-\ket{\mathrm{FM}},
    \quad 
    S^\pm_{\mathbf{q}}\equiv \frac{1}{\sqrt{N}}\sum_{j}e^{i\mathbf{q} \cdot \mathbf{R}_j}S^\pm_j.
    \label{eq:spinWave}
\end{equation}
The energy of the spin wave with wavevector $\mathbf{q}$ is 
\begin{eqnarray}
\label{eq:spin_wave_energies}
    E_\mathbf{q} &=& E_{\mathrm{FM}}+\hbar\omega_\mathbf{q},\\ \label{eq:omega_q}\hbar\omega_\mathbf{q}&=&b+2JS\sum_{\boldsymbol{\delta}} \sin^2\left( \frac{\mathbf{q}\cdot  \boldsymbol{\delta}}{2}\right).
\end{eqnarray}
Here, we have introduced $\{\boldsymbol{\delta}\}$, the set of vectors locating nearest neighbors and we have assumed a Bravais lattice so that there is a nearest neighbor at $-\boldsymbol{\delta}$ for every $\boldsymbol{\delta}$.

The surface electrons are modeled by massive Dirac fermions, with a mass that depends on the state of the impurity spins:
\begin{equation}
    H_{em}=\int d^2r\psi^\dagger(\mathbf{r})\left[-i v_\mathrm{F}\boldsymbol{\sigma}\cdot\boldsymbol{\nabla}+m(\mathbf{r})\sigma_z\right]\psi(\mathbf{r}).
    \label{eq:H_em}
\end{equation}
Here, $\psi(\mathbf{r})=A_\mathrm{surf}^{-1/2}\sum_\mathbf{k}e^{i\mathbf{k}\cdot\mathbf{r}}c_\mathbf{k}$, where $A_\mathrm{surf}$ is the two-dimensional surface area, $c_\mathbf{k}=(c_{\mathbf{k}\uparrow},\, c_{\mathbf{k}\downarrow})^T$, and where $c_{\mathbf{k}s}^\dagger$ creates an electron with wavevector $\mathbf{k}=(k_x,k_y)^T$ and spin $s$. The mass $m(\mathbf{r})$ is an \emph{operator}, which depends on the impurity spins:
\begin{equation}
    m(\mathbf{r})=m_0+\sum_i \mathcal{J}(\mathbf{r}-\mathbf{R}_i)S_i^z,
    \label{eq:m(r)}
\end{equation}
where we have introduced a local finite-ranged exchange coupling $\mathcal{J}(\mathbf{r}-\mathbf{R}_i)$ between each magnetic impurity $i$ at position $\mathbf{R}_i$ and the electron spin density at position $\mathbf{r}$. The constant background mass $m_0$ breaks time-reversal symmetry; it may arise either from a global magnetic field or from some other background magnetization. 

The system Hamiltonian [Eq.~\eqref{eq:H_complete}] includes terms that may arise from an out-of-plane magnetic field ($\propto m_0, \propto b$) through a Zeeman coupling to the electrons, but we neglect the diamagnetic shifts and other orbital effects that would arise from a magnetic vector potential. This is justified for small impurity clusters of size $A_\mathrm{surf}$ and in a sufficiently weak magnetic field $B$: $A_\mathrm{surf}B\ll \Phi_0$, where $\Phi_0=h/e$ is the flux quantum. 
In addition, we neglect terms like $\sim \mathcal{J}_\parallel S^\nu_j\psi^\dagger(\mathbf{r})\sigma_\nu\psi(\mathbf{r})$ [$\nu\in\left\{x,y\right\}$] that couple the in-plane components of the impurity spins to the electron spins. These terms may be ignored in the presence of a large exchange anisotropy, $|\mathcal{J}(\mathbf{r})| \gg |\mathcal{J}_\parallel|$, or for a large effective Zeeman term $|b|\gg \sqrt{2S}|\mathcal{J}_\parallel|$.

In the next section, we show that the elementary excitations of $H_S$ [Eq.~\eqref{eq:H_complete}] have the same form as the magnon states given by Eq.~\eqref{eq:spinWave}. The key difference, as shown below, is that these new excitations come with an electronic state localized to the vicinity of the lowered spin (see Fig.~\ref{fig:fig1}). To create these excitations, starting from the ferromagnetic ground state, in addition to a transverse magnetic field (which creates the spin excitation), we couple our system to an electronic reservoir, which provides the electron that occupies the localized electronic state. This is shown in Sec.~\ref{sec:spin_susceptibility} (see Fig.~\ref{fig:fig3}). 

\section{Jackiw-Rebbi magnons}\label{sec:jackiw_rebbi_magnons}
The total impurity spin in the $z$-direction, $S_{\mathrm{tot}}^z= \sum_iS_i^z$, commutes with $H_S$. The Hamiltonian $H_S$ for the combined electronic and magnetic system can therefore be diagonalized independently in sectors where $S^z_{\mathrm{tot}}$ is fixed. For the ferromagnetic ground state we have $S^z_{\mathrm{tot}}=NS$, giving
\begin{equation}
    \braket{\mathrm{FM}|H_S|\mathrm{FM}}=E_{\mathrm{FM}}+H_{\mathrm{FM}},
\end{equation}
where the effective Hamiltonian $H_{\mathrm{FM}}=\braket{\mathrm{FM}|H_{em}|\mathrm{FM}}$ is obtained from Eq.~\eqref{eq:H_em} with the replacement $m(\mathbf{r})\to\braket{\mathrm{FM}|m(\mathbf{r})|\mathrm{FM}}=m_\mathrm{FM}(\mathbf{r})$. The mass profile $m_{\mathrm{FM}}(\mathbf{r})$ is 
\begin{equation} 
m_\mathrm{FM}(\mathbf{r})=m_0+S\sum_{i}\mathcal{J}(\mathbf{r}-\mathbf{R}_i),
\end{equation} 
and $E_\mathrm{FM}$ is the ferromagnetic ground-state energy, as given in Eq.~(\ref{eq:E_FM}).

Provided that $\mathcal{J}(\mathbf{r})>0$ $\forall\mathbf{r}$ (and if $m_0>0$), the mass $m_{\mathrm{FM}}(\mathbf{r})$ is finite and positive everywhere, leading to a gapped single-particle energy spectrum for $H_{\mathrm{FM}}$. 
The eigenstates of $H_S$, with $S_{\mathrm{tot}}^z=NS$, are then given by a tensor product of electronic and magnetic states; the electronic states are eigenstates of $H_{\mathrm{FM}}$ and the magnetic system is described by the ferromagnetic ground state $\ket{\mathrm{FM}}$.

In the subspace of one spin lowered, $S^z_{\mathrm{tot}}=SN-1$, the magnetic state can be constructed from the basis $\{\ket{\mathbf{R}_i}\}$, $i=1,\hdots,N$, where
\begin{equation}
\ket{\mathbf{R}_i}\equiv \frac{1}{\sqrt{2S}}S_i^-\ket{\mathrm{FM}}.
\end{equation}
Similar to $H_\mathrm{FM}$, we define  $H_i$, the effective Hamiltonian for the electrons when the magnetic system is \textit{frozen} in the state $\ket{\mathbf{R}_i}$:
\begin{equation}
    H_{i}=\braket{\mathbf{R}_i|H_{em}|\mathbf{R}_i},
\end{equation}
which is given by Eq.~\eqref{eq:H_em} with the replacement $m(\mathbf{r})\to m_i(\mathbf{r})=\braket{\mathbf{R}_i|m(\mathbf{r})|\mathbf{R}_i}$:
\begin{equation}
    m_i(\mathbf{r})=m_\mathrm{FM}(\mathbf{r})-\mathcal{J}(\mathbf{r}-\mathbf{R}_i).
    \label{eq:massRi}
\end{equation}
This is the mass term produced by the impurities when the magnetic system is in the state $\ket{\mathbf{R}_i}$ [see Fig.~\ref{fig:fig2}(a)]. 

\subsection{Single-particle Jackiw-Rebbi magnons}

The Hamiltonian $H_i$ has single-particle eigenstates $\ket{\alpha}_i$:
\begin{equation}
    H_i\ket{\alpha}_i=\varepsilon_\alpha\ket{\alpha}_i,
\end{equation}
where the energies $\varepsilon_\alpha$ are independent of $i$ due to the discrete translational invariance of the original Hamiltonian.

We now introduce the composite state made of the $\alpha^{\mathrm{th}}$ electronic eigenstate of $H_i$ and the spin state $\ket{\mathbf{R}_i}$,
\begin{equation}
    \ket{\alpha,\mathbf{R}_i}\equiv\ket{\alpha}_i\otimes \ket{\mathbf{R}_i}.
    \label{eq:composite_state}
\end{equation}
The matrix elements of $H_S$ in this basis are:
\begin{equation}
\begin{split}
    \braket{\alpha,\mathbf{R}_i|H_S|\beta,\mathbf{R}_j}=&\left(E_\mathrm{FM}+b+JSz+\varepsilon_\alpha\right)\delta_{ij}\delta_{\alpha\beta}\\
    &-\gamma_{\alpha\beta}JS\sum_{\boldsymbol{\delta}} \delta_{\mathbf{R}_i,\mathbf{R}_j+\boldsymbol{\delta}}.
    \label{eq:matrixElements}
\end{split}
\end{equation}
Above, we have introduced
\begin{equation}
    \gamma_{\alpha\beta}\equiv \prescript{}{i}{\braket{\alpha| \beta}}_j,\quad i,j\in\braket{ij},
    \label{eq:matrixElements2}
\end{equation}
the electronic overlap between the states $\ket{\alpha}_i$ and $\ket{\beta}_j$. The notation $i,j\in\langle ij\rangle$ indicates that the definition holds for all nearest neighbors $i,j$. For simplicity, we assume a high-symmetry impurity lattice, so that all nearest-neighbor bonds are equivalent. 


\begin{figure}
    \centering
    \includegraphics{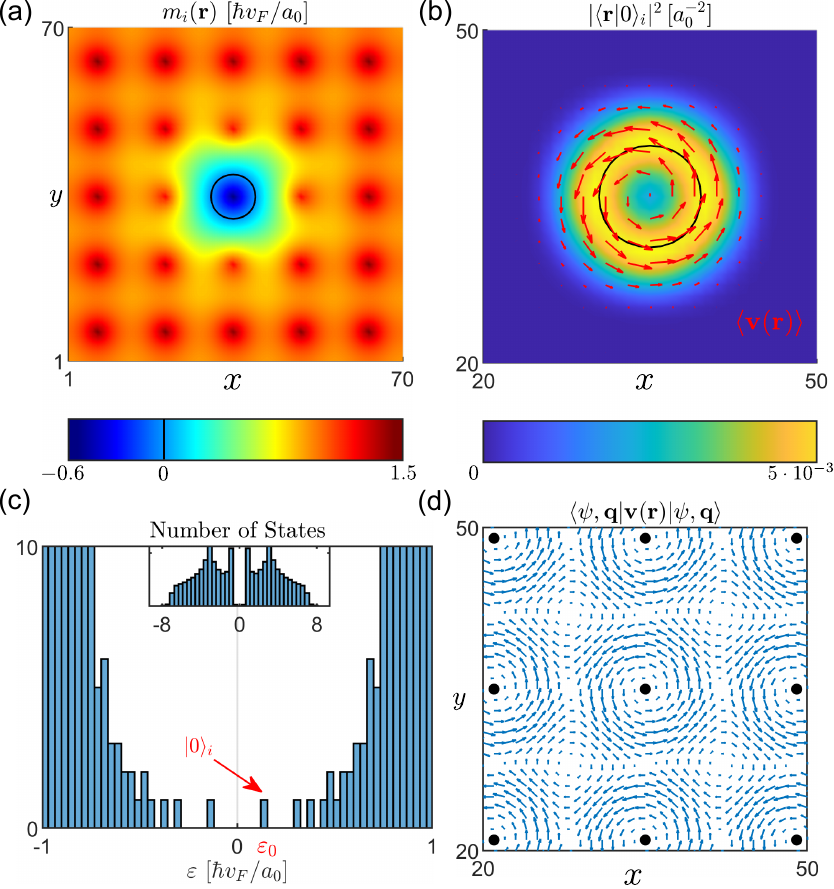}
    \caption{(a) Mass profile, $m_i(\mathbf{r})$, given by Eq.~\eqref{eq:massRi} when $\mathbf{R}_i=(35,35)$. Here, the exchange $\mathcal{J}(\mathbf{r})$ is given by Eq.~\eqref{eq:formofJ}. The black line denotes the region where $m_i(\mathbf{r})=0$, resulting in a sign change in the mass leading to a domain-wall bound state. (b) Probability density (color scale) and local velocity (red arrows) for the lowest positive-energy eigenstate, $\ket{0}_i$, of $H^\mathrm{latt}_i$ [Eq.~\eqref{eq:LattHamiltonian}]. (c) The spectrum of the Hamiltonian $H_{i}^\mathrm{latt}$ is symmetric about $E=0$ as a consequence of particle-hole symmetry.  Isolated  energies are found in the gap, corresponding to bound states, localized to the domain wall. The values for the parameters used are given in Sec.~\ref{sec:numerical}. (d) Velocity profile for either $\ket{\psi,\mathbf{q}}$ or $\ket{\Psi,\mathbf{q}}$, given by Eq.~\eqref{eq:velocity-mbJRM}. The black dots give the positions of magnetic impurities.}
    \label{fig:fig2}
\end{figure}

Jackiw-Rebbi bound states arise if the mass term $m_i(\mathbf{r})$ [Eq.~\eqref{eq:massRi}] change sign in the vicinity of the flipped spin. In the rest of this paper, we focus our attention on the lowest positive-energy single-particle bound state. Although the analysis below also applies to any of the bound states, we concentrate on the lowest positive-energy bound state. This choice is motivated by the many-body case where we begin at half filling and the lowest-energy unoccupied state is then populated.
We label this bound state by $\alpha= 0$, such that $H_i\ket{0}_i=\varepsilon_0\ket{0}_i$ [see Fig.~\ref{fig:fig2}(c)].
The localized Jackiw-Rebbi state $\ket{0,\mathbf{R}_i}$ is coupled to the states $\ket{\alpha,\mathbf{R}_j}$ with coupling strength  $JS\gamma_{0\alpha}$.
If the separation in energy between this state and other electronic states is large relative to the coupling, i.e.
\begin{equation}
    JS\gamma_{0\alpha} \ll  |\varepsilon_0-\varepsilon_\alpha|,\quad (\alpha\ne 0),
    \label{eq:perturbationCondition}
\end{equation}
then we can apply degenerate perturbation theory and diagonalize the Hamiltonian in the subspace spanned by the set of degenerate states $\{\ket{0,\mathbf{R}_1},\hdots,\ket{0,\mathbf{R}_{N}}\}$. In this subspace, the effective Hamiltonian is equivalent to a tight-binding model with nearest-neighbor hopping proportional to the overlap $\gamma\equiv\gamma_{00}$ [see Eqs.~(\ref{eq:matrixElements}-\ref{eq:matrixElements2})]. The eigenstates of this model are plane waves of the form
\begin{equation}
    \ket{\psi,\mathbf{q}}=\frac{1}{\sqrt{N}}
    \sum_{j}e^{i\mathbf{q}\cdot\mathbf{R}_j}\ket{0,\mathbf{R}_j},
    \label{eq:JRM}
\end{equation}
which we call Jackiw Rebbi Magnons (JRMs). These states differ from conventional spin waves; in addition to carrying a single quantum of angular momentum (spin), they carry a single charge associated with a bound electron. Because these states are derived from a tight-binding model, they inherit a simple tight-binding dispersion:
\begin{equation}
    \tilde{E}_\mathbf{q}=
    E_{\mathrm{FM}}+\hbar\tilde{\omega}_\mathbf{q},
\end{equation}
with
\begin{equation} \label{eq:omega_tilde_q}
    \hbar\tilde{\omega}_\mathbf{q}=\varepsilon_0+b+JSz(1-\gamma)+2JS\gamma \sum_{\boldsymbol{\delta}} \sin^2\left(\frac{\mathbf{q}\cdot \boldsymbol{\delta}}{2}\right).
\end{equation}

\subsection{Many-body Jackiw-Rebbi magnons}

The analysis above is not restricted to single-particle states. For example, the half-filled ($N_e$-particle) state resulting from filling the negative-energy single-particle eigenstates of $H_i$ is
\begin{equation}
    \ket{\mathrm{hf},\mathbf{R}_i} \equiv \ket{\mathrm{hf}}_i \otimes  \ket{\mathbf{R}_i};\quad\ket{\mathrm{hf}}_i=\prod_{\alpha\,(\varepsilon_\alpha<0)} c_{\alpha i}^\dagger \ket{\mathrm{vac}}_e,
    \label{eq:hf_i}
\end{equation}
where $c_{\alpha i}^\dagger\ket{\mathrm{vac}}_e=\ket{\alpha}_i$ with $\ket{\mathrm{vac}}_e$ the electronic vacuum.
The minimum-energy $N_e$-conserving excitation out of the half-filled state has a finite energy cost $\Delta\varepsilon=2\varepsilon_{0}$  (the spectrum of $H_i$ is symmetric around zero because it is particle-hole symmetric \cite{chiu2016classification}). 
A general $N_e$-particle eigenstate can be written as
\begin{equation}
\begin{split}
    \ket{\boldsymbol{\alpha}_{N_e},\mathbf{R}_j}&\equiv \ket{\boldsymbol{\alpha}_{N_e}}_j\otimes \ket{\mathbf{R}_j},  \\\ket{\boldsymbol{\alpha}_{N_e}}_j&= \left(c^\dagger_{\alpha_1j}\cdots c^\dagger_{\alpha_{N_e}j}\right)\ket{\mathrm{vac}}_e.
\end{split}
\end{equation} 
If the minimum excitation energy, $2\varepsilon_0$, is large compared to the matrix element that connects the states $\ket{\mathrm{hf},\mathbf{R}_i}$ with $\ket{\boldsymbol{\alpha}_{N_e},\mathbf{R}_j}$, that is
\begin{equation}
    JS\prescript{}{i}{\braket{\mathrm{hf}|\boldsymbol{\alpha}_{N_e}}}_j \ll 2\varepsilon_0,\quad i,j\in \braket{ij},
\end{equation}
then we can diagonalize the Hamiltonian in the degenerate subspace consisting of $\{\ket{\mathrm{hf},\mathbf{R}_1},\hdots,\ket{\mathrm{hf},\mathbf{R}_{N}}\}$ to obtain the states
\begin{equation}
    \ket{\mathrm{hf},\mathbf{q}}\equiv \frac{1}{\sqrt{N}}\sum_je^{i\mathbf{q}\cdot \mathbf{R}_j}\ket{\mathrm{hf},\mathbf{R}_j}.
    \label{eq:hf_q}
\end{equation}
Using the same argument, $H_S$ has approximate $(N_e+1)$-particle eigenstates (many-body JRMs) of the form
\begin{equation}
    \ket{\Psi,\mathbf{q}}\equiv\frac{1}{\sqrt{N}}\sum_je^{i\mathbf{q}\cdot \mathbf{R}_j}c_{0j}^\dagger \ket{\mathrm{hf},\mathbf{R}_j}.
    \label{eq:manyBodyJRM}
\end{equation}
We use the term \emph{many-body} Jackiw-Rebbi magnons specifically for these $(N_e+1)$-particle states which, in addition to carrying a quantum of angular momentum (spin), also carry an additional electronic charge beyond half filling. To differentiate this many-body state from the single-particle state $\ket{\psi,\mathbf{q}}$ we use the capital letter $\Psi$. Note that the hopping between $(N_e+1)$-particle states with a single spin-flip, $c_{0j}^\dagger \ket{\mathrm{hf},\mathbf{R}_j}$, is now proportional to the many-body overlap 
\begin{equation}
    \gamma \to \gamma_\Psi =\braket{\mathbf{R}_i,\mathrm{hf}| c_{0i}c^\dagger_{0j}|\mathrm{hf},\mathbf{R}_j}, (i,j)\in \langle ij\rangle,    
\end{equation}
 which may decrease rapidly as the number of particles increases. 

\subsection{Numerical Example}\label{sec:numerical}

As a concrete example, we consider a square-lattice model that recovers $H_i$ in the long-wavelength limit. This allows us to find lattice versions of the eigenstates $\ket{0}_i$. 
The lattice model is obtained by first rewriting the kinetic-energy term in Eq.~\eqref{eq:H_em} in $\mathbf{k}$-space:
\begin{equation}
    v_\mathrm{F}\int d^2r\psi^\dagger(\mathbf{r})\left[-i \boldsymbol{\sigma}\cdot\boldsymbol{\nabla}\right]\psi(\mathbf{r})  = v_\mathrm{F} \sum_{\mathbf{k}}c_{\mathbf{k}}^\dagger \mathbf{k}\cdot\boldsymbol{\sigma}c_{\mathbf{k}},
\end{equation}
then substituting $k_\nu\rightarrow a_0^{-1}\sin k_\nu a_0,\nu\in\left\{x,y\right\}$. To maintain a single Dirac cone (in the limit $m_0\to0,\,\mathcal{J}(\mathbf{r})\to 0$) at the $\Gamma$-point and to avoid extra Dirac points at the Brillouin-zone corners, we furthermore introduce an additional term $-2A\sigma_z(2-\cos k_x a_0-\cos k_y a_0)$. The equivalent term in a continuum model is $-Aa_0^2\sigma_zk^2$ \cite{shen2012topological}. 
The resulting real-space lattice Hamiltonian is therefore
\begin{equation}
\begin{split}
    H_i^\mathrm{latt} =& \sum_{\mathbf{r}}
    c^\dagger_{\mathbf{r}} m_i(\mathbf{r}) \sigma_z c_{\mathbf{r}}
    \\&+\frac{i}{2}\frac{\hbar v_\mathrm{F}}{a_0} \sum_{\mathbf{r},\nu}\left(
    c^\dagger_{\mathbf{r}+a_0\mathbf{e}_\nu} \sigma_\nu c_{\mathbf{r}}
    - \mathrm{h.c.} \right)
    \\&+ 
    A\sum_{\mathbf{r},\nu} \left( c^\dagger_{\mathbf{r}+a_0\mathbf{e}_\nu} \sigma_z c_{\mathbf{r}} +\mathrm{h.c.}-2c_\mathbf{r}^\dagger\sigma_zc_\mathbf{r}\right),
\end{split}
\label{eq:LattHamiltonian}
\end{equation}
where $\nu\in\left\{x,y\right\}$.

For the simulations (Fig.~\ref{fig:fig2}) we consider a $5\times 5$ impurity grid with impurity spacing $a=14a_0$ on top of a crystalline lattice of size $N_x=N_y=70$, where $N_\nu$ is the number of lattice sites in the $\nu$-direction and consider periodic boundary conditions. Although here we consider a square lattice of impurity spins for this particular numerical example, note that the results obtained in Sec.~III apply for any choice of Bravais lattice. We use $A/(\hbar v_\mathrm{F}/a_0)=1, m_0/(\hbar v_\mathrm{F}/a_0)=-1/2,S=1/2$. For $\mathcal{J}(\mathbf{r}-\mathbf{R}_i)$, we use the exponentially decaying function
\begin{equation}
    \mathcal{J}(\mathbf{r}-\mathbf{R}_i)=\mathcal{J}_0e^{-|\mathbf{r}-\mathbf{R}_i|/\lambda},
    \label{eq:formofJ}
\end{equation}
with $\mathcal{J}_0/(\hbar v_\mathrm{F}/a_0)=2,\lambda=a/2$, and set $v_\mathrm{F}=a_0=1$. 
The parameters chosen here lead to a low-energy bound state encircling the flipped impurity spin associated with a circulating (chiral) current. The specific choice of parameters taken above is not necessary to obtain this bound state. The only requirement is that there is a gap closure in the vicinity of the flipped impurity spin. In the continuum model, this requirement amounts to a change in sign of $m_i(\mathbf{r})$, while in the lattice model this requires a change in sign of $m_i(\mathbf{r})[m_i(\mathbf{r})-8A]$ in the vicinity of the lowered impurity (see Eq.~(3.91) in Ref.~\cite{shen2012topological}). Even when the eigenstates vary on a scale comparable to the lattice constant (far from the long-wavelength limit), the qualitative physics of electronic states bound to magnetic domain walls will be preserved.

Figure~\ref{fig:fig2}(a) shows the spatially dependent Dirac mass $m_i(\bm r)$ [Eq.~\eqref{eq:massRi}] resulting from a flipped impurity spin. Figure \ref{fig:fig2}(b) shows the lowest-energy Jackiw-Rebbi-like bound state in the vicinity of the flipped impurity spin. In Fig~\ref{fig:fig2}(c) the electronic density of states of $H_i^\mathrm{latt}$ is shown, where bound states are clearly visible within the gap. The number and energies of these in-gap bound states vary with the size of the negative-mass domain \cite{hurst2015charged}. The local velocity for $\ket{0}_i$ and $\ket{\psi,\mathbf{q}}$ are shown in Figs.~\ref{fig:fig2}(b) and~\ref{fig:fig2}(d), respectively. 
The local velocity is evaluated from the Heisenberg equation of motion,
\begin{equation}
    \mathbf{v}(t)\equiv\dot{\hat{\mathbf{r}}}(t)=\frac{i}{\hbar}[H_i^\mathrm{latt},\hat{\mathbf{r}}(t)],
\end{equation}
where $\hat{\mathbf{r}}=\sum_{\mathbf{r}}\mathbf{r}c^\dagger_\mathbf{r} \sigma_0 c_\mathbf{r}$ is the position operator.
Therefore, the local velocity operator in the $\nu$-direction is 
\begin{equation}
\begin{split}
    v_\nu(\mathbf{r})=\left(\frac{v_\mathrm{F}}{2}
    c^\dagger_{\mathbf{r}+\mathbf{e}_\nu}\sigma_\nu c_{\mathbf{r}}-iA\frac{a_0}{\hbar} c^\dagger_{\mathbf{r}+\mathbf{e}_\nu}
    \sigma_z c_{\mathbf{r}} +\mathrm{h.c.}
    \right).
\end{split}
\end{equation}

The JRM [Eq.~\eqref{eq:JRM}] and the many-body JRM [Eq.~\eqref{eq:manyBodyJRM}] have the same velocity profile since
\begin{equation}\label{eq:velocity-mbJRM}
\begin{split}
    \braket{\Psi,\mathbf{q}|\mathbf{v}(\mathbf{r})|\Psi,\mathbf{q}}&=\frac{1}{N}\sum_i \braket{\mathrm{hf},\mathbf{R}_i|c_{0i}\mathbf{v}(\mathbf{r})c^\dagger_{0i}|\mathrm{hf},\mathbf{R}_i}\\&=
    \frac{1}{N}\sum_i \prescript{}{i}{\braket{0|\mathbf{v}(\mathbf{r})|0}}_i.
\end{split}
\end{equation}
The off-diagonal elements in the expression above, $\braket{\mathrm{hf},\mathbf{R}_i|\mathbf{v}(\mathbf{r})|\mathrm{hf},\mathbf{R}_{j\ne i}}$, vanish because $v_\nu(\mathbf{r})$ only acts on the electronic space. The average velocity is then independent of the momentum $\mathbf{q}$ of the Jackiw-Rebbi magnon. Moreover, the contribution from the negative-energy states vanishes because the Hamiltonian $H_i$ has particle-hole symmetry and the velocity operator is odd under the particle-hole transformation. 

Including the orbital effect of a vector potential can break the  particle-hole symmetry and thus alter the calculation above. However, a weak out-of-plane magnetic field will preserve the existence of in-gap domain-wall bound states \cite{hurst2015charged}, leading to a similar qualitative picture to that presented here.

\section{Spin Susceptibility}\label{sec:spin_susceptibility}

\begin{figure}
    \centering
    \includegraphics{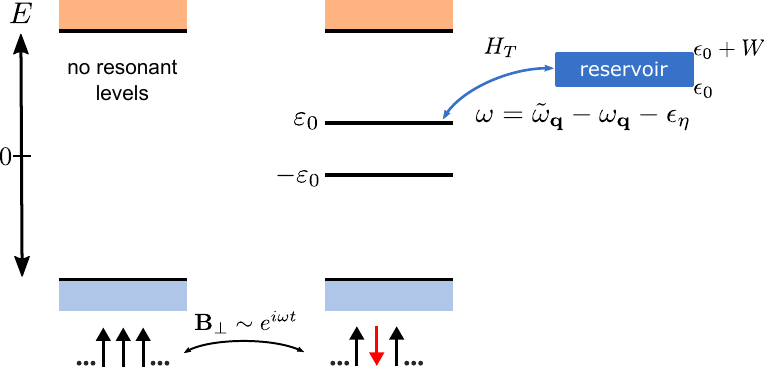}
    \caption{The electronic states available when the magnetic system is in the ferromagnetic ground state $\ket{\mathrm{FM}}$ are not resonant with the reservoir. The energy range of the reservoir is such that the spin-wave state, with excitation energy $\omega_\mathbf{q}$ [Eq.~\eqref{eq:omega_q}], may only be resonant with the JRMs, with energy $\tilde{\omega}_\mathbf{q}-\epsilon_\eta$ [Eq~\eqref{eq:omega_tilde_q}] (here, $\epsilon_\eta$ is the energy of the reservoir electron that is removed to populate the Jackiw-Rebbi state). The JRM excitations are only accessible after a spin is flipped (magnon is created) through a transverse magnetic field $\mathbf{B}_\perp(t)$. The tunneling Hamiltonian $H_T$ couples the electronic reservoir with the surface states of the 3D topological insulator [see Eq.~\eqref{eq:H_T}].}
    \label{fig:fig3}
\end{figure}

The dynamic transverse spin susceptibility can be used to probe the existence of JRMs. In contrast with the case of a Heisenberg Hamiltonian [Eq.~\eqref{eq:HeisenbergHam}], for which a time-varying transverse magnetic field directly excites purely magnetic excitations (magnons), the transverse magnetic field alone is not sufficient to excite JRMs, which also carry a bound electronic charge. For this reason, in addition to a time-varying transverse magnetic field, here we also incorporate an electronic reservoir (see Fig.~\ref{fig:fig3}).  

The Hamiltonian of our system (the topological-insulator surface electrons and magnetic impurities) subjected to a transverse magnetic field and coupled to an electronic reservoir is given by
\begin{eqnarray}
    H(t) & = & H_0+V(t),\\
    H_0 & = & H_S+H_R+H_T,
\end{eqnarray}
where $V(t)\propto\sum_{j}\mathbf{B}_\perp(\mathbf{R}_j,t)\cdot \mathbf{S}_j$ contains the spatially and temporally varying transverse magnetic field $\mathbf{B}_\perp=\left(B_x,B_y,0\right)^T$ (we neglect the electron-field coupling as we consider the case in which the impurity-field coupling dominates), $H_S$ is defined in Eq.~\eqref{eq:H_complete}, $H_R$ describes the electronic reservoir, and $H_T$ describes tunneling to the reservoir:
\begin{eqnarray}
H_R & = & \sum_\eta\epsilon_\eta a_\eta^\dagger a_\eta,\\
H_T&=&\sum_{\eta,j} \left( t_{\eta j} c_{0j}^\dagger a_\eta +\mathrm{h.c.}\right)\ket{\mathbf{R}_j}\bra{\mathbf{R}_j}.\label{eq:H_T}
\end{eqnarray}
The operator $a^\dagger_\eta$ creates an electron with energy $\epsilon_\eta$ in the reservoir, where $\eta$ is a general quantum number. We have included only those terms in $H_T$ that couple to the relevant mid-gap bound states [Fig. \ref{fig:fig3}(a)] under the assumption that all other levels are far from resonance. In general, the tunnel coupling $t_{\eta j}$ associated with $c_{0j}^\dagger$ will depend on the impurity $j$, but for simplicity here we take $t_{\eta j}=t_{\eta}$ (independent of $j$), which can be realized when the extended reservoir states couple equivalently to all impurity sites. This choice guarantees that $H_T$ preserves the wavevector $\mathbf{q}$ in converting a spin wave to a JRM: $_e\bra{\mathrm{vac}}\bra{\mathbf{q}}H_T\ket{\psi,\mathbf{q}'}\propto \delta_{\mathbf{q},\mathbf{q}'}$. 
 
The zero-temperature dynamic transverse susceptibility is
\begin{equation}
    \chi^{+-}(\mathbf{q},\omega)=\int dt e^{i\omega t -0^+|t|}C(\mathbf{q},t).
\end{equation}
Here, the spin-spin correlation function is given by
\begin{equation}
    C(\mathbf{q},t)=-i\theta(t)\braket{\mathrm{gs}|[\hat{S}_\mathbf{q}^+(t),S^-_{-\mathbf{q}}]|\mathrm{gs}},
    \label{eq:transvSusc}
\end{equation}
where $\hat{S}^+_\mathbf{q}(t)=e^{iH_0t}S_\mathbf{q}^+e^{-iH_0t}$ is the time-dependent raising operator in the interaction picture and $\ket{\mathrm{gs}}$ is the intially prepared ground state of $H_0$. 

We now consider a special initial condition that allows for an exact solution. In particular, we consider the fully spin-polarized ferromagnetic ground state, a filled band of reservoir states, but no electrons occupying any of the surface states of the topological insulator. Such an initial condition is an eigenstate of the effective Hamiltonian $H_0$, but in practice it would only be metastable, eventually collapsing due to non-resonant tunneling terms that have been neglected in $H_T$. We discuss the more realistic case of a filled surface-state valence band, below. Explicitly, the assumed ground state is:
\begin{equation}\label{eq:gs_empty}
    \ket{\mathrm{gs}}=\ket{\mathrm{vac}}_e \otimes \ket{\mathrm{FM}} \otimes \ket{\mathrm{FB}}_R,
\end{equation}
where $\ket{\mathrm{FB}}_R$ represents the filled reservoir band.
For an initially filled reservoir, the tunneling Hamiltonian only leads to transitions within a restricted subspace of states, spanned by
\begin{eqnarray}\label{eq:short_notation_1}
\ket{\psi(0)} & = & \frac{1}{\sqrt{2S}} S_{-\mathbf{q}}^-\ket{\mathrm{gs}},\\\label{eq:short_notation_2}
\ket{\eta} & = & \ket{\psi,\mathbf{-q}}\otimes a_\eta\ket{\mathrm{FB}}_R.
\end{eqnarray}
Since $H_T$ preserves the wavevector $\mathbf{q}$, we have suppressed the explicit $\mathbf{q}$-dependence in $\left\{\ket{\psi(0)},\,\ket{\eta}\right\}$, above.

We can directly solve the time-dependent Schr\"odinger equation for this problem,
\begin{equation}
    i\partial_t \ket{\psi(t)}=H_0\ket{\psi(t)},
\end{equation}
giving a solution (see Refs.~\onlinecite{gurvitz2017wave,gaveau1995limited} for equivalent solutions):
\begin{equation}
    \ket{\psi{(t)}} = \psi_0(t)\ket{\psi(0)}+\sum_\eta \psi_\eta(t)\ket{\eta}.
\end{equation}
This solution then gives the correlation function $C(\mathbf{q},t)$ (and hence, the susceptibility): 
\begin{equation}
    C(\mathbf{q},t) = -i\theta(t)e^{iE_\mathrm{gs}t}2S\braket{\psi(0)|\psi(t)}.
\end{equation}
Here, the ground state energy (defined by $H_0\ket{\mathrm{gs}}=E_\mathrm{gs}\ket{\mathrm{gs}}$) is
\begin{equation}\label{eq:gs_energy}
    E_\mathrm{gs}\equiv E_\mathrm{FM}+\sum_\eta \epsilon_\eta.
\end{equation}
The exact susceptibility is then
\begin{equation}
\label{eq:susceptibility}
    \chi^{+-}(\mathbf{q},\omega)=\frac{2S}{\omega-\omega_\mathbf{q}-\Sigma_\mathbf{q}(\omega)},
\end{equation}
where the self-energy is
\begin{equation} \label{eq:self_energy_term}
    \Sigma_\mathbf{q}(\omega)=\sum_\nu \frac{|t_\eta^2|}{\omega-\left(\tilde{\omega}_\mathbf{q}-\epsilon_\eta\right)+i0^+}.
\end{equation}
In the equations above, $\omega_\mathbf{q}$ and $\tilde{\omega}_\mathbf{q}-\epsilon_\eta$ correspond to the excitation energy for a spin wave and a JRM with momentum $\mathbf{q}$, respectively [cf. Eqs.~\eqref{eq:omega_q} and~\eqref{eq:omega_tilde_q}]. We additionally introduce phenomenological decay rates for the JRMs ($\Gamma_\mathrm{JRM}$) and for the magnons ($\Gamma_m$) with the replacements
\begin{eqnarray}
    0^+ & \to & \Gamma_\mathrm{JRM},\\
    \Sigma_\mathbf{q}(\omega) & \to & \Sigma_\mathbf{q}(\omega)-i\Gamma_m.
\end{eqnarray}

\begin{figure}
    \centering
    \includegraphics{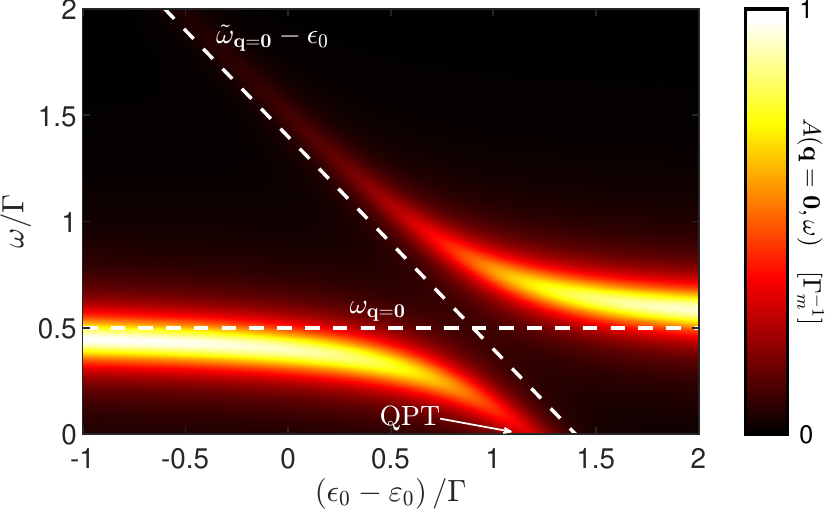}
    \caption{Spectral function for a narrow-band reservoir ($W\to 0$) as a function of the reservoir energy $\epsilon_0$. The two peaks correspond to magnon-JRM polaritons. A quantum phase transition (QPT) occurs when the excitation energy for the lowest-energy magnon-JRM polariton becomes negative. In this plot we have chosen the parameters $\Gamma=1,S=1/2,b=1/2,J=1/2,\varepsilon_0=5,\gamma=1/10,\Gamma_m=\Gamma_\mathrm{JRM}=1/10$, where all energies are given in units of $\Gamma$.}
    \label{fig:fig4}
\end{figure}

Evidence of JRMs can be found by analyzing peaks of the spectral function,
\begin{equation}
    A(\mathbf{q},\omega)=-2\,\mathrm{Im}\,\chi^{+-}(\mathbf{q},\omega).
\end{equation}
Due to the tunneling term $H_T$, neither the spin-wave $\ket{\psi(0)}\propto S_{-\mathbf{q}}\ket{\mathrm{gs}}$, nor the JRM with a reservoir hole at energy $\epsilon_\eta$, $\ket{\eta}$, is an eigenstate of $H_0$. As a result, $A(\mathbf{q},\omega)$ shows multiple peaks associated with the excitation energies for hybridized magnon-JRM polaritons (see the example illustrated in Fig.~\ref{fig:fig4}). The characteristic double-peaked structure and avoided crossing is distinct from the case of an isolated Heisenberg ferromagnet, having susceptibility
\begin{equation}\label{eq:chi_0}
    \chi_0^{+-}(\mathbf{q},\omega)=\frac{2S}{\omega-\omega_\mathbf{q}+i0^+}.
\end{equation}
Here, the corresponding spectral function $A_0(\mathbf{q},\omega)=-2\,\mathrm{Im}\, \chi_0^{+-}(\mathbf{q},\omega)$ is given by a Dirac delta distribution centered at the spin-wave excitation energy $\omega_\mathbf{q}$: $A_0(\mathbf{q},\omega)\propto \delta(\omega-\omega_\mathbf{q})$. 

As an illustrative example, we consider the reservoir band to form a quasi-continuum of energies $\epsilon_\eta\in\left[\epsilon_0,\epsilon_0+W\right]$, as shown in Fig.~\ref{fig:fig3}(a). We additionally assume a constant density of states $\rho$ and constant tunnel coupling $t_\eta=t$. In the narrow-band limit, $W\rightarrow0$, the self-energy is then
\begin{equation}
    \Sigma_\mathbf{q}(\mathbf{q},\omega)=\frac{\Gamma/\pi}{\omega-(\tilde{\omega}_\mathbf{q}-\epsilon_0)+i0^+},
\end{equation}
where $\Gamma=\pi\rho|t|^2$. In this simple limit, the susceptibility has only two poles corresponding to two magnon-JRM polaritons. The magnon-like resonance with frequency $\sim\omega_\mathbf{q}$ couples to the JRM-like resonance having frequency $\sim \tilde{\omega}_\mathbf{q}-\epsilon_0$, with an avoided crossing of width $\sim \Gamma$ (Fig.~\ref{fig:fig4}).

A quantum phase transition (QPT) takes place if the lowest-energy magnon-JRM polariton falls in energy below the ferromagnetic ground state. In the case illustrated in Fig.~\ref{fig:fig4}, this corresponds to
\begin{equation}
    \omega_\mathbf{q}(\tilde{\omega}_\mathbf{q}-\epsilon_0)<\left(\Gamma/\pi \right)^2.
\end{equation}
Above a critical value of the reservoir energy $\epsilon_0>\epsilon_\mathrm{crit}=\tilde{\omega}_\mathbf{q}-\Gamma^2/(\omega_\mathbf{q}\pi^2)$, the fully spin-polarized ground state becomes unstable to the formation of JRMs. The observation of such a transition due to a coupled reservoir could be a robust indicator of the formation of JRMs. This transition can also be seen in the spectral function (Fig.~\ref{fig:fig4}), where the lower magnon-JRM polariton peak moves to negative frequency at large $\epsilon_0$.

In Eq.~\eqref{eq:gs_empty} we considered an initial ground state with empty bands for the topological-insulator surface, but a filled band of reservoir states. Such an initial condition is only metastable if the reservoir electrons have a finite amplitude for tunneling to the surface. A more realistic (but computationally more complex) situation is therefore realized when the topological insulator surface is at half-filling, corresponding to a global chemical potential positioned just above the top of the reservoir band ($\mu=\epsilon_0+W+0^+$ in Fig.~\ref{fig:fig3}). In this case, the ground state is given by
\begin{equation}
    \ket{\mathrm{hf},\mathrm{FM}}\equiv\left(\prod_{v\,(\varepsilon_v<0)}c_v^\dagger\right)\ket{\mathrm{vac}}_e\otimes\ket{\mathrm{FM}}\otimes\ket{\mathrm{FB}}_R.
    \label{eq:MB_gs}
\end{equation}
Here, $c_v^\dagger$ creates an electron in a single-particle eigenstate of $H_\mathrm{FM}$. To calculate the susceptibility, we compute the action of $S_{-\mathbf{q}}^-$ on this state:
\begin{equation}
\begin{split}
    S_{-\mathbf{q}}^-&\ket{\mathrm{hf},\mathrm{FM}}=\\&\sum_j e^{-i\mathbf{q}\cdot \mathbf{R}_j}\left(\prod_{v\,(\varepsilon_v<0)}c_v^\dagger\right)\ket{\mathrm{vac}}_e\otimes\ket{\mathbf{R}_j}\otimes\ket{\mathrm{FB}}_R.
    \label{eq:Sq_MB_gs}
\end{split}
\end{equation}
Unlike the case of the initial state $\ket{\mathrm{gs}}$, here the action of $S_{-\mathbf{q}}^-$ does not simply generate a magnon excitation independent of the state of the surface electrons. Instead, to rewrite Eq.~\eqref{eq:Sq_MB_gs} in terms of $H_S$ eigenstates, the creation operators $c_v^\dagger$ should be expanded, for each $j$, in terms of electronic states that account for the position of the spin flip:
\begin{equation}
    c_v^\dagger=\sum_{\alpha} U_{v,\alpha j} c_{\alpha j}^\dagger,
\end{equation}
with some coefficients $U_{v,\alpha j}$. 

Since the single-particle electronic orbitals must adapt to the magnetic configuration, the character of the half-filled state will also change, depending on the site of the spin flip. In addition to producing the half-filled state for each $j$, the spin flip will generally scatter surface electrons to many excited states, resulting in a complex dynamics, different from the case described above for the initial state $\ket{\mathrm{gs}}$. However, the results obtained for the initial state $\ket{\mathrm{gs}}$ could still provide insight into the qualitative behavior for the half-filled ground state considered above, provided there is a mechanism for rapid equilibration of the surface electrons. To illustrate this point, consider the $N_e$-electron state
\begin{equation}
    \ket{\mathrm{hf}}_\mathrm{FM}\equiv \left(\prod_{v(\varepsilon_v<0)}c_v^\dagger \right)\ket{\mathrm{vac}}_e. 
\end{equation}
Rewriting this in the eigenbasis of $H_j$:
\begin{equation}
    \ket{\mathrm{hf}}_\mathrm{FM}=\prescript{}{j}{\braket{\mathrm{hf}|\mathrm{hf}}}_\mathrm{FM}\ket{\mathrm{hf}}_j+\sum_{\boldsymbol{\alpha}} \prescript{}{j}{\braket{\boldsymbol{\alpha}_{N_e}|\mathrm{hf}}}_\mathrm{FM} \ket{\boldsymbol{\alpha}_{N_e}}_j.
\end{equation}
The lowest-energy $N_e$-particle eigenstate of $H_j$ is $\ket{\mathrm{hf}}_j$ [see Eq.~\eqref{eq:hf_i}]. If the surface electrons interact with a dissipative environment (e.g. phonons) that quickly relaxes the electronic state $\ket{\mathrm{hf}}_\mathrm{FM}$ to $\ket{\mathrm{hf}}_j$ in Eq.~\eqref{eq:Sq_MB_gs}, the state $S_{-\mathbf{q}}^-\ket{\mathrm{hf},\mathrm{FM}}$ will relax into the state $\ket{\mathrm{hf},-\mathbf{q}}\otimes\ket{\mathrm{FB}}_R$ [see Eq.~\eqref{eq:hf_q}]. This eigenstate of $H_S+H_R$ couples only to the continuum of many-body JRMs $\{\ket{\Psi,-\mathbf{q}}\otimes a_\nu\ket{\mathrm{FB}}_R\}$, similar to the empty-band case considered above. We can then obtain the susceptibility, which will have the same qualitative behavior as that given in Eq.~\eqref{eq:susceptibility}. There would, however, be corrections to the short-time dynamics of the spin-spin correlation function due to relaxation effects, modifying the spectral function at high frequency. A precise microscopic accounting of this relaxation mechanism and the associated effect on the spectral function is, however, beyond the scope of this work.

\section{Conclusions}\label{sec:Conclusions}
In this paper, we have introduced a model that can describe Dirac electrons on the surface of a 3DTI coupled to magnetic impurities, which themselves interact via a ferromagnetic Heisenberg exchange. In contrast to previous work, where the magnetic system was treated classically \cite{liu2009magnetic,yu2010quantized,hurst2015charged,Kunal2017magnetoconductance,Kunal2019TunableSkyrmion,Tokura2019,tserkovnyak2012thin,efimkin2021topological}, here we have considered a quantum magnet, which together with the Dirac electron give rise to new excitations - the Jackiw-Rebbi-Magnons (JRMs). We have systematically derived the form of these excitations in two situations: First, we have solved a single-particle problem, in which the conduction and valence bands of the 3DTI surface are initially empty, and second, we have solved  a many-body problem, starting from a half-filled 3DTI surface. While the low-energy theory is well captured by a continuum description of the Dirac electrons, we have also constructed a lattice model with the appropriate low energy limit. This lattice model allowed us to numerically study the density of electronic states, the creation of chiral domain-wall bound states, and the associated current and magnetic textures for JRMs (shown in Fig.~\ref{fig:fig2}). Finally, we studied the dynamic transverse spin susceptibility.   In order to probe the JRM states, we coupled our system to an electronic reservoir, allowing a charge to be transferred to the surface, thus producing a JRM that can then contribute to the magnetic susceptibility.  In contrast with the conventional magnon response, which would be independent of the reservoir energy, we find a response that disperses with the reservoir energy due to the formation of magnon-JRM polaritons (Fig.~\ref{fig:fig4}). In fact, this model shows a quantum phase transition at a critical value of the reservoir energy, where the fully spin-polarized ground state will collapse in favor of the production of magnon-JRM polaritons.

This paper is limited to the study of single  JRMs that carry a single unit of spin angular momentum and a single unit of charge. An interesting direction for future work would be to consider an alternative class of JRMs involving multiple spin flips, but still binding a single charge. Alternatively, it would be interesting to consider multiply charged JRMs. In each case, new interaction effects should be taken into account and these effects may lead to a rich family of excitations. In addition to the conventional magnon-magnon interactions present in a Heisenberg ferromagnet, JRMs produced from multiple spin flips would be influenced by the molecular binding energy of a delocalized bound electron, similar to the situation that may arise for classical magnetic skyrmions on the surface of a 3DTI \cite{Kunal2019TunableSkyrmion}. For multi-electron JRMs (or for multiple JRMs, each with a single electronic charge), it will be important to account for the Coulomb interaction, in addition to magnetic interactions and the molecular binding energy.
Furthermore, we only consider a specific type of magnetic system: a Heisenberg ferromagnet in a Bravais lattice with nearest-neighbor interactions. There is, however, a rich family of magnetic systems where electronic Jackiw-Rebbi states could bind to magnons, including ferromagnetic materials with next-nearest-neighbor interactions, systems with multiple single-magnon bands coming from sublattice degrees of freedom (on, e.g., a honeycomb lattice), and antiferromagnetic materials. Extending the present analysis to this broader range of magnetic systems would be another interesting direction for future investigation.

\begin{acknowledgments}
IM-B acknowledges a fellowship from Consejo Nacional de Ciencia y Tecnolog\'ia (CONACYT) and Mitacs. TP-B and WAC acknowledge support from Natural Sciences and Engineering Research Council of Canada (NSERC), Institut Transdisciplinaire d'Information Quantique (INTRIQ), and Fonds de Recherche du Qu\'ebec--Nature et Technologies (FRQNT).
\end{acknowledgments}


\bibliography{JRMagnons.bib}

\end{document}